# COMBINED HYDROTHERMAL LIQUEFACTION OF POLYURETHANE AND LIGNOCELLULOSIC BIOMASS FOR IMPROVED CARBON RECOVERY

Juliano Souza dos Passos [a,b], Stefano Chiaberge [c], Patrick Biller* [a,b]

[a] Department of Biological and Chemical Engineering, Aarhus University, Hangøvej 2, DK-8200 Aarhus N, Denmark
[b] Aarhus University Centre for Circular Bioeconomy, DK-8830 Tjele, Denmark
[c] Eni s.p.a. Renewable Energy & Environmental R&D, via Fauser 4, 28100, Novara, Italy.
* Corresponding author

**Abstract.**

Due to the high versatility of Polyurethane (PUR), its share amongst synthetic polymers to manufacture consumer goods is increasing. This study proposes, tests and validates through pilot processing a highly efficient method for carbon recovery of PUR and in the form of an oil phase concentrated in carbon using hydrothermal liquefaction (HTL). Hot liquid water mixed with PUR residues and a lignocellulosic material (two species of Miscanthus) were treated at subcritical water temperatures, generating an oil-phase rich in hydrocarbons. A high synergistic effect in the co-liquefaction was observed, leading to a carbon and chemical energy recovery to the oil of 71 and 75% respectively. Pilot plant processing, using optimized process parameters, yielded a total process efficiency, accounting heating utilities, of 61%, resulting in a 3.2 ratio of energy return over investment. Using spectroscopic and high-resolution mass spectrometry analysis the oil revealed a high content of nitrogen hetero aromatic and polyol compounds. The high synergy observed for the co-liquefaction of PUR and miscanthus is attributed to recombination of synthetic and biological materials, specifically due to aromatics present in the media and nitrogen-containing compounds recombination. The results show that HTL can be an efficient method for carbon recovery of PUR aided by biomass.

## 1. INTRODUCTION

Polyurethanes (PUR) are a group of extremely versatile synthetic polymers. Their applications range from automobile seats, chairs, sofas, footwear and carpets to refrigerators, insulation boards, medical applications, coatings, binders and many others.[1] Such a variety of applications and continuous innovations have resulted in a PUR yearly production of around 26.4 Mton worldwide, with a production growth rate of 6.25% per year [2013-2018].[1,2] PUR is usually divided into flexible and rigid foams, coatings, adhesives, sealants and elastomers.[1,3] The main difference between these groups are the polyols' composition and ratios of isocyanides and polyols used in production. PUR manufacturing plants are known to be wasteful, generating a considerable amount of trims and scrap, reaching 10% of the total PUR production.[2,3] Despite the fact that part of the waste material produced has recycling opportunities that yield secondary products

(carpets, inferior quality insulation boards), the amount diverted to such applications does not alleviate waste handling, particularly for post-consumer products[2], and hence, innovation is still needed for PUR recycling.

The variety of PUR applications also brings challenges upon general recycling methods that can cope with all kinds of inputs. The most common recycling methods are mechanical (regrinding, rebinding, adhesive pressing, injection moulding and compression moulding), mainly yielding lower quality products than the initial feedstock source. These processes can cope with mixed flexible PUR, though the preference is usually for pure inputs. Nevertheless, mechanical recycling cannot handle rigid PUR. All of the mechanical recycling methods, besides adhesive pressing, have limitations regarding recycled feedstock, as it tends to decrease product quality.[2,3]

Chemical recycling methods for PUR are known and can be divided into pyrolysis, gasification, hydrogenation and solvolysis (hydrolysis, aminolysis, glycolysis). Pyrolysis typically produces a considerable amount of HCN, NO, CO in the gas phase[2], together with volatiles that include methane, ethylene, benzene and others.[4] The process usually yields (at >450 °C) 5–25%$_{wt}$ char, 10–45%$_{wt}$ liquids and >40%$_{wt}$ gases. Due to the undesired products present in the gas phase, pyrolysis is not preferred when dealing with PUR. A recent catalytic gasification report shows that hydrogen rich gases (up to 80% hydrogen fraction) are achievable, with gas yields around 70% at 1100 °C.[5] Gasification also entails the issue of generating pollutants, such as NO, HCN, and other nitrogen containing gases. Hydrogenation on the other hand, as it is conducted in a reductive media, tends to produce $NH_3$ from the feedstock nitrogen. This process generates comparable amounts of volatiles as pyrolysis, with greater energy content due to the hydrogen saturation towards hydrocarbons.[2,6]

The most used and developed solvolysis method for PUR recycling is glycolysis with existing examples of pilot plants and industrial scale facilities.[3] As the glycol reactant attaches to urethane bonds via transesterification reactions, the polyols from the feedstock are released and become available for further use.[3] Several catalysts have been tested for this approach, however amines are typically employed, though their presence in products is undesired.[2] Many research groups and patents have described PUR glycolysis employing different glycols as reactants for different PURs and a range of catalysts, generally with high recovery of polyols, sometimes with two phase separation.[2,3,7] Despite this method being the most developed, with pilot and industrial scale plants around the world, the complexity of the products obtained restrict their use to rigid foam manufacturing at limited contribution to virgin feedstock.[3]

Hydrolysis of PUR is usually described employing dry superheated steam at mild temperatures and low pressures (190-250 °C, 10-20 bar) for relatively long residence times (>30 min).[3] Products obtained are complex and difficult to separate, hindering process development so far to laboratory and pilot scale. We have described before that hydrothermal liquefaction (HTL) using 350 °C liquid water successfully depolymerizes PUR, generating an oil rich in polyols and amines free from urethane bonds.[8] HTL is a comparable process to hydrolysis, however, HTL has the advantage of not using steam as a water source – i.e. avoiding evaporation energy spent on steam generation –, and recovering products in a separated oil phase. Typically, HTL processing has been developed for dealing with organic wastes – e.g. sewage sludge[9–11], manure[12], food waste[13] –, lignocellulosic[14] and algae[15] . Currently, several pilot plants are operational[14,16] and the technology readiness level of HTL is considered to be 5.

The combined HTL (co-HTL) of synthetic polymers and different materials has been suggested before, for instance on reports about microalgae co-HTL with polypropylene[17]; macroalgae with polyethylene, polypropylene and polyamides[18]; lignocellulosic materials with polypropylene, polyethylene, PET and polyamides[19] and; more recently catalytic processing of lignocellulosic materials and polypropylene.[20] All studies find synergies and antagonistic effects for selected mixtures and materials, which indicates co-HTL is an interesting approach for processing polymers together with biomasses. Recently, we have reported that *Miscanthus Giganteus* – a model for lignocellulosic biomass – presents significant synergy effects in co-HTL with PUR foam.[21] Those findings encouraged the present study, where we develop a predictive model for oil, carbon and energy yields based on batch experiments of PUR and *Miscanthus Giganteus*, validating the models in continuous processing using a continuous flow pilot plant.[14] We also further describe the synthetic oil's detailed composition employing Fourier-Transform Ion Cyclotron Resonance Mass Spectrometry (FTICR-MS) analysis, and use this information to describe possible reaction mechanisms that yield the observed synergistic effects on the basis of the molecular level characterization.

## 2. MATERIALS

*Miscanthus Giganteus* (used in batch experiments) and *Miscanthus Lutarioriparius* (used in continuous experiments), hereby referred to as M, were harvested at Aarhus University's facilities at Foulum, Denmark. PUR foam utilized in batch experiments was taken from a disposed chair, while the PUR foam utilized in continuous experiments was kindly provided by Dan-Foam ApS (Tempur Denmark). Both types of PUR and M utilized in batch and continuous experiments were characterized via TGA, FTIR and compositional analysis (comparison available in Figure S1, Figure S2 and Table S2). When comparing *Miscanthus Giganteus* and *Miscanthus Lutarioriparius,* a higher fixed carbon content was found present in the former, shown in the TGA measurements depicted in Figure S1. This difference likely arises from lignin content variations between the two species. Nevertheless, the other characterizations (Figure S2 and Table S2) are considered sufficiently similar for direct comparison of results.

## 3. METHODS

3.1. **Batch HTL**

Batch experiments were conducted using 20 mL bomb-type reactors and procedure already described.[22] Table S1 depicts the set of experiments conducted in batch mode to screen blends of M and PUR from pure M to pure polyurethane in steps of 25% change within the temperature range of 300 and 350 °C. In each experiment, an appropriate amount of each material summing 1.00 g of dry matter was added to the reactor together with 8.00 g of water. After sealing, the reactors were inserted in a fluidized pre-heated sand bath at the desired temperature. After reaction time, reactors were immediately quenched in water to room temperature. Gases were vented, water decanted and the solids and oil mixture separated via solvent-assistant filtration (Methanol). Oil mass was weighted after solvent evaporation. Gas, solids and oil yields are based on initial feed weight, according to Equation 1.

Equation 1
$$Yield_{component} = \frac{m_{component}}{m_{feed}}$$

Predictive models for the oil, carbon and energy yields were derived from the dataset (acquired according to Table S1) using Equation 2 regression.[13] As mixtures tested were binary, i.e. $PUR + M = 1$ (where PUR is the dry basis polyurethane concentration in feed and M is the dry ash free basis M concentration in feed), the model was shortened for a single variable on feed material concentration, M, for simplification. Temperature (T) is expressed in °C.

Equation 2
$$Yield\ (T, M) = a_o + \sum_{i=1}^{2} b_i\ T^i + \sum_{i=1}^{3} c_i\ M^i + \sum_{i=1;j=1}^{i=2;j=3} d_{i,j}\ T^i\ M^j$$

The Bayesian information criterion (BIC) was used to exclude terms in an iterative manner, simplifying the model without losing significance. Analysis of variance (ANOVA) was also conducted for the final predictive models.

3.2. **Elemental analysis**

CHNS-O (oxygen by difference) was determined for solid and oil fractions of both batch and continuous experiments using an Elementar vario Macro Cube elemental analyser (Langenselbold, Germany). The Channiwala-Parikh correlation (Equation 3) was used to estimate the HHV for oils. Energy and carbon yields were calculated using Equation 4 and Equation 5, respectively.

Equation 3
$$HHV\ \left[\frac{MJ}{kg}\right] = 0.3491\ C + 1.1783\ H + 0.1005\ S - 0.1034\ O - 0.0151\ N - 0.0211\ A$$

Equation 4
$$Energy\ yield\ (\eta_{th})\ \% = \frac{HHV_{oil}\ \left[\frac{MJ}{kg_{oil}}\right] \cdot Yield_{oil}\ \left[\frac{kg_{oil}}{kg_{feed}}\right]}{HHV_{feed}\ \left[\frac{MJ}{kg_{feed}}\right]} \times 100$$

Equation 5
$$Carbon\ yield\ \% = \frac{C_{oil}\left[\frac{kg_C}{kg_{oil}}\right].Yield_{oil}\left[\frac{kg_{oil}}{kg_{feed}}\right]}{C_{feed}\left[\frac{kg_C}{kg_{feed}}\right]} \times 100$$

## 3.3. Attenuated Total Reflectance – Fourier Transformed Infra-Red (ATR-FTIR)

Selected solid and oil samples were analyzed using a Bruker Alpha Platinum ATR FTIR spectrometer (24 spectra collected from 4000 to 400 cm$^{-1}$) with resolution of 2 cm$^{-1}$. In between analyzes, 96% ethanol was used for cleaning the diamond crystal before baseline collection. Solid samples were compressed against the crystal and oil samples were rubbed on top for measurements.

## 3.4. Thermogravimetric analysis (TGA)

A TGA Mettler Toledo SDTA851 was used to analyze raw materials, oil and solid products. The TGA was operated using constant heating rate of 10 K min$^{-1}$ from 50 °C to 900 °C under nitrogen followed by 10 minutes under air at constant temperature. A minimum of 5 mg of mass sample was placed in the TGA ceramic crucibles.

## 3.5. Fourier Transform Ion Cyclotron Resonance Mass Spectrometry (FTICR MS)

Bio-Crude samples were diluted with a mixture of chloroform and acetonitrile (1:20) to a final concentration of 0.4 mg/ml. Mass spectrometry analysis were performed on a 7T FTICR MS (LTQ-FT Ultra Thermo Scientific), equipped with atmospheric pressure chemical ionization (APCI) ion source. The mass spectra were collected in positive ion mode. The samples were infused at a flow rate of 100 uL min$^{-1}$ using the following typical APCI (+) conditions: source heater 380°C, source voltage 5 kV, capillary voltage 7 V, tube lens voltage 60 V, capillary temperature at 275 °C, sheath gas 60 arbitrary units, auxiliary gas 10 arbitrary units. The mass spectra were acquired in positive mode with a mass range of m/z 100-1400. The resolution was set to 200,000 (at m/z 400). The ion accumulation time was defined by the automatic gain control (AGC), which was set to 10$^6$. 360 scans were acquired for each analysis to improve the signal to noise ratio using the Booster Elite system (Spectroswiss), that allowed to directly register the transient data. Transients were then processed by the software Peak by Peak-Petroleomic version (Spectroswiss). The 360 transients were averaged and then Fourier Transformed into a single averaged mass spectrum. The resulting spectrum was further processed to remove noise (thresholding set to 6 Sigma of the background noise), and internally recalibrated through the unwrapping method.[23] Around 10,000 different peaks were then obtained. Final attribution of these peaks was conducted using the composing function of the software Peak by Peak with the error limit of ±2ppm.

Molecular formulas were categorized according to different parameters, such as the number of heteroatoms (N and O) and number of unsaturations expressed as DBE (Double Bond Equivalents).[24] For each molecular formula the DBE was calculated according to Equation 6 (for $C_cH_hN_nO_oS_s$).

$$\text{Equation 6} \quad DBE = c - \frac{h}{2} + \frac{n}{2} + 1$$

Classes of compounds were assigned according to the heteroatoms present in each molecule, and their relative abundances were used for building class distribution plots. The more abundant classes were then plotted in double bond equivalent (DBE) versus carbon number plots according to their carbon number, DBE value and relative abundance in the mass spectrum.

### 3.6. Continuous HTL

PUR waste and raw M were milled using a modified twin screw extruder (Xinda, 65 mm twin screw extruder with 2000 mm barrel length) and mixed in the optimal proportion identified from the batch experiments. Fresh water and carboxymethyl cellulose (0.5% total slurry weight) were added to prepare the slurry using a 2 m$^3$ paddle mixer together with a Microcut MCH-D 60 A wet mill to ensure homogeneity and pumpability. The slurry was fed in the continuous HTL pilot plant located at Aarhus University, Foulum, described by our group previously.[14] Figure 1 depicts the temperature profile over the pilot plant during the production campaign. The reactor section temperature was kept at 316.4 ± 6.2 °C and slurry feed mass flow was 55.1 ± 0.6 kg/h. The campaign ran for 5h and results presented are based in thermal steady state operation, as highlighted in Figure S3. The thermal steady state was verified using the logarithmic mean temperature difference in the heat exchanger. Steady state was determined when this value changes was less than 5%, which was achieved after 2h of continuous operation for a total steady state period of approximately 3h (see Figure S3).

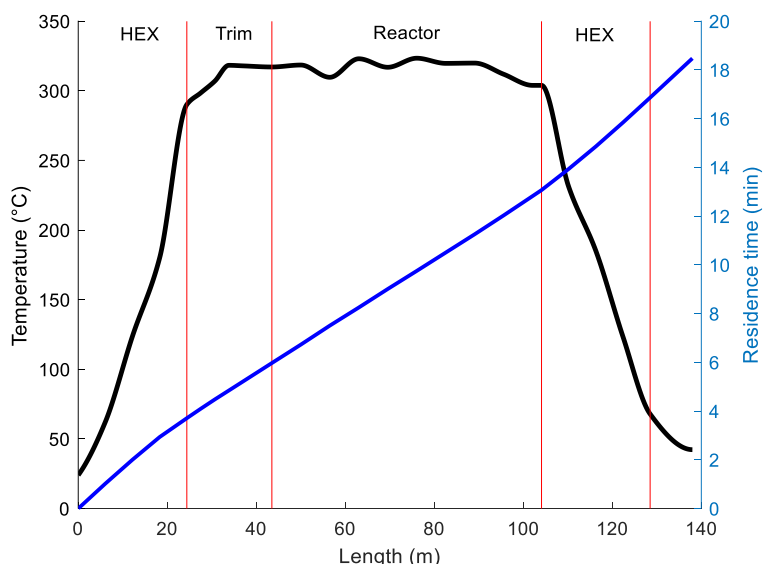

Figure 1 - Pilot plant temperature profile during campaign

The synthetic oil product was separated gravimetrically from water and 10 g raw oil aliquots were diluted in methanol and vacuum filtered to determine solid content. Water content in the raw oil was determined via Karl Fischer titration. The total energy efficiency ($\eta_{tot}$) and the energy return over investment (EROI) for continuous processing were calculated using Equation 7 and Equation 8, respectively.[14]

Equation 7
$$\eta_{tot} = \frac{E_{Oil}\ [kW_{oil}]}{(E_{feed} + E_{pump} + E_{Trim\ heat} + E_{Reactor\ heat})\ [kW_{total\ input}]} \times 100$$

Equation 8
$$EROI = \frac{E_{Oil}\ [kW_{oil}]}{(E_{pump} + E_{Trim\ heat} + E_{Reactor\ heat})\ [kW_{external\ input}]}$$

## 4. RESULTS AND DISCUSSION

### 4.1. Batch HTL and yield predictive models

Equation 9, Equation 10 and Equation 11 show the oil, carbon and energy yield predictive models, respectively, derived from Equation 4 and Equation 5 results regressed for Equation 2 and simplified stepwise using the BIC. ANOVA correspondent to these models are found in Table S3, Table S4 and Table S5, respectively for oil, carbon and energy yields. Statistically significant p-values still present in the models were allowed to comply with lower BIC, enhancing prediction efficiency. Other statistical tests were performed to test the model for bias and are depicted in Figure S4, Figure S5 and Figure S6 for oil, carbon and energy yield models respectively. The data shows all models being highly capable of predicting yields with relatively low bias. Generally, the empirical terms presented are within the same orders of magnitude.

Equation 9
$$Oil_{yield} = -7.3701 + 5.2221\ 10^{-2}\ T + 10.712\ M - 8.6216\ 10^{-5}\ T^2 - 7.1146\ 10^{-2}\ T\ M + 0.15073\ M^2 + 3.4747\ 10^{-5}\ T^2\ M - 1.2577\ 10^{-2}\ T\ M^2 + 2.2617\ M^3$$

Equation 10
$$Carbon_{yield} = -1.9371 + 1.8474\ 10^{-2}\ T - 2.1977\ M + 1.270\ 10^{-2}\ T\ M - 3.421\ 10^{-5}\ T^2 - 1.3397\ M^2 - 9.5083\ 10^{-3}\ T\ M^2 + 2.4806\ M^3$$

Equation 11
$$Energy_{yield} = 1.7772 - 4.0843\ 10^{-3}\ T - 2.5976\ M + 1.4308\ 10^{-2}\ T\ M - 1.0344\ M^2 - 1.1120\ 10^{-2}\ T\ M^2 + 2.5693\ M^3$$

**Error! Reference source not found.** A, B and C show respectively the oil, carbon and energy yields heat maps based on respective models (Equation 9, Equation 10 and Equation 11). It is possible to identify the maximum oil yield with 65.7% at 314 °C and M 0.225 $g_{miscanthus}/g_{DM}$. Temperatures between 300 and 325 °C and M between 0.175 and 0.30 $g_{miscanthus}/g_{DM}$ all depict oil yields higher than 64%. When compared to pure feedstock materials (PUR and M), where oil yields range from 35 to 45%, there is a remarkable 30-40% increase. Within the range tested, effects of M are more pronounced than temperature, as can be observed in **Error! Reference source not found.**, indicating that lower temperatures may be used as long as feedstock mixture is optimized for maximum oil production. The carbon yield peaks at 70.8% at 308 °C and M 0.25, while the energy yield maximum reaches 75.0% in the same M, but at 300 °C. Overall, temperatures between 300 and 325 should be preferred, while M mixtures in PUR of 0.20 to 0.28 yield best results in all three parameters mapped.

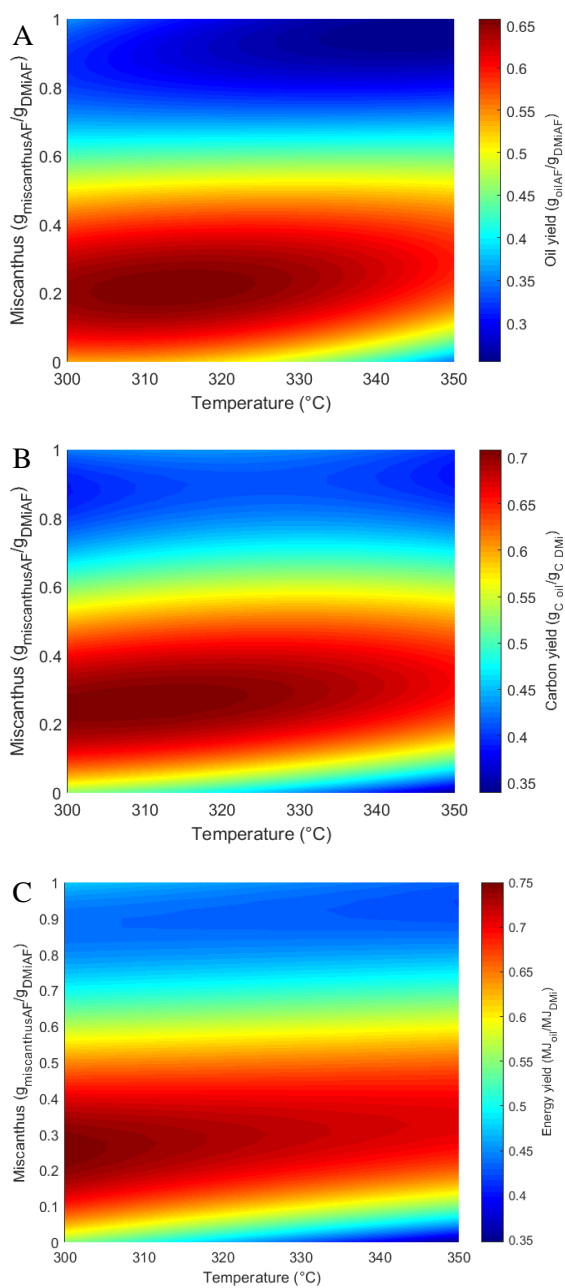

Figure 2 – Oil (A), carbon (B) and energy (C) yields prediction model heatmap

The surfaces response depicted in **Error! Reference source not found.** A, B and C are clearly similar, confirming that the superior oil yield carries carbon and chemical energy to this product fraction. Usually HTL of common biomass-derived feedstock materials do not reach such high efficiencies for oil yields, particularly when not using lipid-rich biomaterials or oil recycling strategies, yields of up to 50% have been reported [16,25]. For example, co-HTL synergies of lignocellulosic materials and sewage sludge are reported to increase oil yields to up to 47.1%[10], while co-HTL of lignocellulosic biomass and manure can reach 30%.[26] The synergistic effect previously reported of *Miscanthus Giganteus* and PUR[21] was significantly improved from 43% oil, 52% carbon and 55% energy yields at 50:50 mix ratio and 350 °C to the aforementioned optimal conditions upon reaction parameters optimization. This indicates a specific stoichiometric relation is present, which will be clarified further upon oil composition discussion. Controlling the presence of lignocellulosic materials and PUR in HTL media can clearly enhance chemical recycling opportunities of PUR.

In literature, batch bomb-type reactor models have been developed to investigate HTL because of their characteristic fast heating rates, which are very similar to the heating profiles observed in continuous HTL plants.[27,28] In the following section, the models described here, based on batch reactor results, are validated using a continuous HTL reactor.

## 4.2. Continuous HTL

The feedstock preparation for the continuous campaign, using the ratio of M and PUR identified in the predictive models was successful and pumpability was achieved. This was not the case for PUR/water slurries alone, thus the addition of lignocellulosic biomass to PUR also facilitates the formation of a pumpable slurry. The resulting slurry exhibited clear signs of non-Newtonian behavior by syringe tests performed in-situ. The slurry was successfully pumped using helical rotor pumps and a Graco Check-Mate positive displacement pump. No clogs nor significant pressure difference over the pilot plant components (heat exchanger hot and cold sections, reactor and take-off system) were observed during the entire run, which indicates the slurry is stable upon pressurized heating and the products do not tend to crosslink.

Table 1 depicts a summary of the continuous campaign results, including feed slurry mass and energetic characterization over the analyzed timeframe of steady state operation, together with the products aqueous phase and biocrude and the energy utilities. The mass flow rate was around 55.1 kg.h$^{-1}$, giving a total residence time of approximately 18.5 min (Figure 1). This mass flow carried into the reactor 32.1 kg$_{Carbon}$.h$^{-1}$, of which 71.2 % was found in the oil phase, 9.8% in the aqueous phase and, by difference, the remaining 19% were converted either to gas or solid products. These results are superior when compared to pure miscanthus HTL, high-nitrogen biomasses such as *Spirulina* or sewage sludge using the same pilot plant.[14]

The nitrogen balance shows that out of the 2.46 kg$_N$.h$^{-1}$ fed to the reactor, 57.9% were found in the oil phase, which contained 3.95% nitrogen by weight, while 16.2% were collected in the aqueous phase, which contained 1123 mg$_N$.L$^{-1}$. Even though there is a negative contribution of N to the HHV, the oil phase still presented a relatively high value of 31.1 MJ.kg$^{-1}$. The nitrogen content in oil is comparable to biomass-derived biocrudes, even though its yield is higher.[9,14,29] The total nitrogen in the aqueous phase and its yield to this fraction are lower than biomass HTL processing results, especially compared to manures and sewage sludge.[30,31]

The chemical energy fed to the pilot plant was 57.5 kW, which was heated using 7.9 kW in the trim heater and 5 kW in the reactor section. The oil product carried 43.1 kW of energy, which is 75.0% of the initial feedstock slurry. Accounting the utilities, the total energy efficiency is 60.7%, with an energy return over investment (EROI) of 3.2. This data leads the PUR/M co-HTL to be an attractive strategy to recover carbon from used materials in an oil form.

Table 1 – Continuous production campaign data summary and results.

| Property | Value | ± | Unit |
|---|---|---|---|
| *Feed slurry* | | | |
| Miscanthus concentration in feed | 0.22 | - | $g_{miscanthus}/g_{feed}$ |
| PUR concentration in feed | 0.78 | - | $g_{PUR}/g_{feed}$ |
| PUR : M ratio in feed | 3.52 | - | $g_{PUR}/g_{miscanthus}$ |
| Mass flow rate | 55.1 | 0.6 | kg/h |
| Dry matter content | 13.77% | 0.01 | (wt.%) |
| Energy in Feedstock | 57.5 | 0.2 | ($kW_{Feed}$, dry) |
| *Product aqueous phase* | | | |
| Total Organic Carbon | 8824 | 48 | mg / L |
| Total Inorganic Carbon | 20 | 2 | mg / L |
| Aqueous phase carbon yield | 9.8% | - | ($kg_{C\,AP}/kg_{C\,Input}$) |
| Total Nitrogen | 1123 | 4 | mg / L |
| Aqueous phase nitrogen yield | 16.2% | - | ($kg_{N\,oil}/kg_{N\,input}$) |
| *Product biocrude* | | | |
| Bio-crude yield (dry solids free) | 65.6% | - | ($kg_{oil}/kg_{input}$) |
| Bio-crude carbon content (dry basis) | 63.1% | 0.11 | ($kg_C/kg_{oil}$) |
| Bio-crude carbon yield (dry solids free) | 71.2% | - | ($kg_{C\,oil}/kg_{C\,input}$) |
| Bio-crude Nitrogen content (dry basis) | 3.95% | 0.14 | ($kg_N/kg_{oil}$) |
| Bio-crude Nitrogen yield (dry solids free) | 57.9% | - | ($kg_{N\,oil}/kg_{N\,input}$) |
| HHV bio-crude (dry solids free) | 31.1 | 0.5 | (MJ/$kg_{oil}$) |
| Energy in Bio-crude (dry solids free) | 43.1 | 0.2 | ($kW_{Oil}$, dry) |
| *External energy input* | | | |
| Trim heater energy requirements | 7.9 | 0.4 | (kW) |
| Reactor energy requirement | 5.0 | 0.0 | (kW) |
| Main pump energy requirement | 0.48 | - | (kW) |
| *Energy efficiency* | | | |
| Energy yield ($\eta_{th}$) | 75.0% | - | ($kW_{Oil}/kW_{Feed}$) |
| Total efficiency ($\eta_{tot}$) | 60.7% | 0.2 | ($kW_{Oil}/kW_{total\,input}$) |
| EROI | 3.2 | 0.2 | ($kW_{Oil}/kW_{external\,input}$) |

### 4.2.1. Thermogravimetric Analysis (TGA)

The batch sample acquired at 325 °C, with mixing ratio of 75:25 (PUR : M) was selected as the most representative in comparison to the already mentioned sample collected from the continuous campaign. Figure 3 compares TGA and DTG results from batch oils of M, PUR and co-HTL, together with the continuous oil campaign. It is clear that the co-HTL oil, either from batch or continuous, is much more similar to the PUR oil than the M biocrude. The co-HTL oils and the PUR one all have a sharp main weight loss around 400 °C. This DTG peak is slightly above the 360-390 °C main decomposition peak found in the raw PURs (Figure S2), which is related to the polyalcohols' thermal degradation.[4,32,33] Both batch and continuous co-HTL oils have a different behavior at temperatures < 400 °C. The former loses around 8% of its weight, while the latter reaches 20% weight loss, both yielding minor broad peaks at 200 °C and 280 °C respectively. After the 400 °C peak, the batch co-HTL oil still holds around 15% of its initial weight, while the continuous campaign oil behaves similarly to the pure batch PUR, both with < 5%. This could be caused due to the higher amount of lignin in the M species used in batch (indicated via TGA solid residue amount in Figure S2).

Overall the TGA and DTG results point that the oil derived from co-HTL of PUR and M is more similar to the pure PUR oil than to pure M. The findings are partially surprising, given the M addition increases significantly the oil phase recovery. However, the amount of PUR participation in the initial mixtures is indeed higher and so the characteristics measured by TGA and DTG follow.

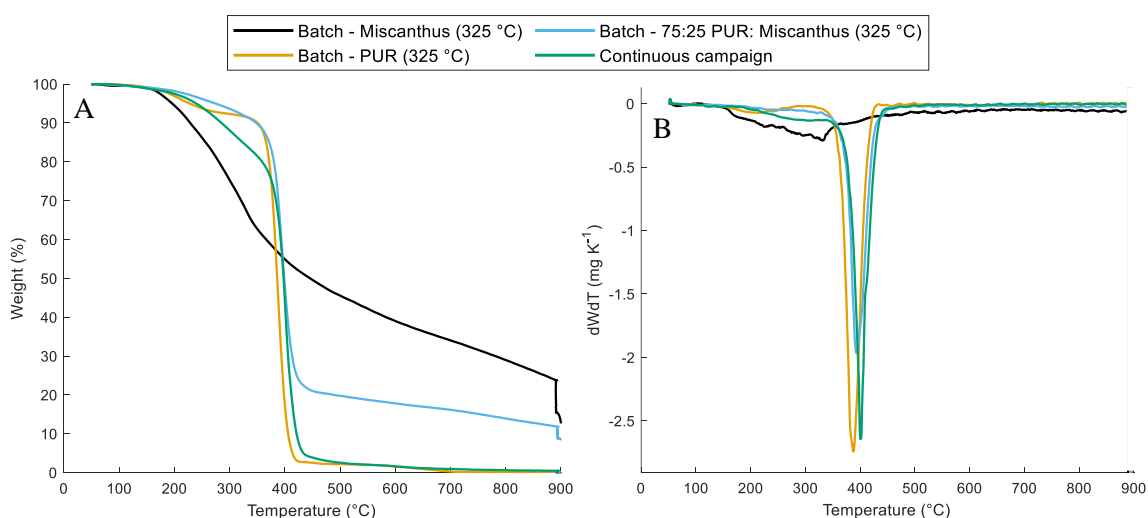

Figure 3 – TGA and DTG of oil from HTL experiments in batch and continuous. Batches: *Miscanthus Giganteus*, PUR, 75:25 (PUR : M), 325 °C.

### 4.2.2. ATR-FTIR

Figure 4 depicts an ATR-FTIR comparison of batch biocrudes from *Miscanthus Giganteus*, PUR, one sample of 75:25 (PUR:M, 325 °C) and the continuous campaign biocrude (filtered using methanol, water free). The M biocrude presents its typical peaks related to phenolics, aliphatics, acids, ketones and alcohols.[34,35] Hydroxyl groups are located between 3030-3660 cm$^{-1}$ with a broad peak, while C–O bonds can also be identified at 1110 and 1205 cm$^{-1}$ and C=O at 1690 cm$^{-1}$. Characteristic C–H bonds can be verified at 2928 (–CH$_2$), 2846 (–CH$_3$), 1460-1350 cm$^{-1}$ and 740-830 cm$^{-1}$. Lastly, C=C are shown in 1603 and 1512 cm$^{-1}$.

The PUR oil FTIR spectra depicted in Figure 4 shares significant resemblance to the raw PUR (Figure S2). However the oil does not contains the characteristic urethane bonds at 1223 (C(=O)O) and 1533 (C–N) cm$^{-1}$, rather presenting a clear N-H peak at 1624 cm$^{-1}$ and a much broader peak in the 3030-3600 cm$^{-1}$ region, related to OH presence together with primary and secondary amines. It is worth noticing a shift from 3291 to 3358 cm$^{-1}$ (from raw feedstock to oil) peak, indicating a change from secondary to primary amine. Other peaks, present in both the oil and the raw material are ether C–O (1090 cm$^{-1}$), double bonded C=C (1590-1630 cm$^{-1}$) and characteristic C–H bonds around 2800 – 3000 cm$^{-1}$.

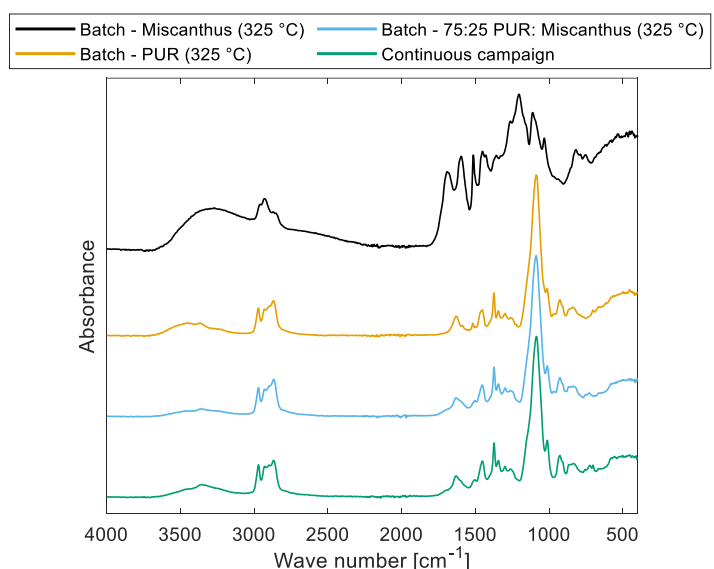

Figure 4 – ATR-FTIR of oil from HTL experiments in batch and continuous. Batches: *Miscanthus Giganteus*, PUR, 75:25 (PUR : M 325 °C).

It is clear the biocrude yielded from the 75:25 (PUR:M, 325 °C) batch is more similar to the PUR oil than the M biocrude. The only differences observed are the broad peaks between 3030-3660 cm$^{-1}$ related to OH and NH groups and, also a broadening of the 1624 cm$^{-1}$ N–H peak. When comparing both oils, it is possible to see the PUR oil has a wider 3030-3660 cm$^{-1}$ peak, indicating a higher occurrence of OH groups. The peak at 3358 cm$^{-1}$, connected to the presence of primary amines can be observed in both oils spectra. The continuous campaign and batch (75:25 PUR:M, 325 °C) oils do not present differences, indicating they share very similar chemical groups and structures in comparable proportion.

### 4.2.3. FTICR MS analysis

The FTICR mass spectra contains over ten thousands different molecular ion peaks (mainly protonated molecular ions), thus some data representation choices were made to clarify the composition, including the classification of elemental formulas of the detected ions based on N and O numbers, C number, DBE, H/C, O/C and N/C ratios.

Figure S7 shows the DBE versus C number for all classes detected in the M batch oil (325 °C). No compounds with nitrogen were detected, as shown in Figure 5, thus families were divided only according to oxygen number (which were identified from 2 to 7). The results depict an oil that contains molecules with 8 to 68 carbons, with relatively more presence from $C_5$ to $C_{40}$, generally increasing average C number directly proportional to oxygen number. The DBE values of the most intense species are between 5 and 20, therefore, most of the compounds likely belong to aromatic oxygen containing species. Despite not detecting nitrogen containing compounds, the analysis presented here is in good agreement to other lignocellulosic materials' HTL biocrudes characterized using similar techniques.[36,37]

Figure 5 depicts the PUR batch (325 °C) oil compound families according to their abundance. It is possible to observe that the most prominent abundances are located in Figure 5A, where only compounds without nitrogen are located. A bell-shaped curve is seen, peaking around family $O_{15}$, tailing towards lower O numbers. These compounds can be best observed in Figure 6A, where the carbon number is plotted over oxygen for all DBE compounds found. This observation is very characteristic of oligomers or polymers. In this case, they are clearly derived from the polyol group of PUR resins.[32,33] Thus the PUR oil has a main fraction characterized by the water insoluble fraction of the polyols present in the original feedstock. The remaining identified compounds have either 1, 2, 4 or 6 N atoms (predominantly 4) accompanied by up to 3 O atoms, as shown in Figure S10. Overall the synthetic oil derived from PUR alone differs greatly from nitrogen rich biomass-derived biocrudes, which may present families with up to 4 N atoms accompanied by up to 3 O atoms, especially with regards to relative abundance and DBE.[31,37]

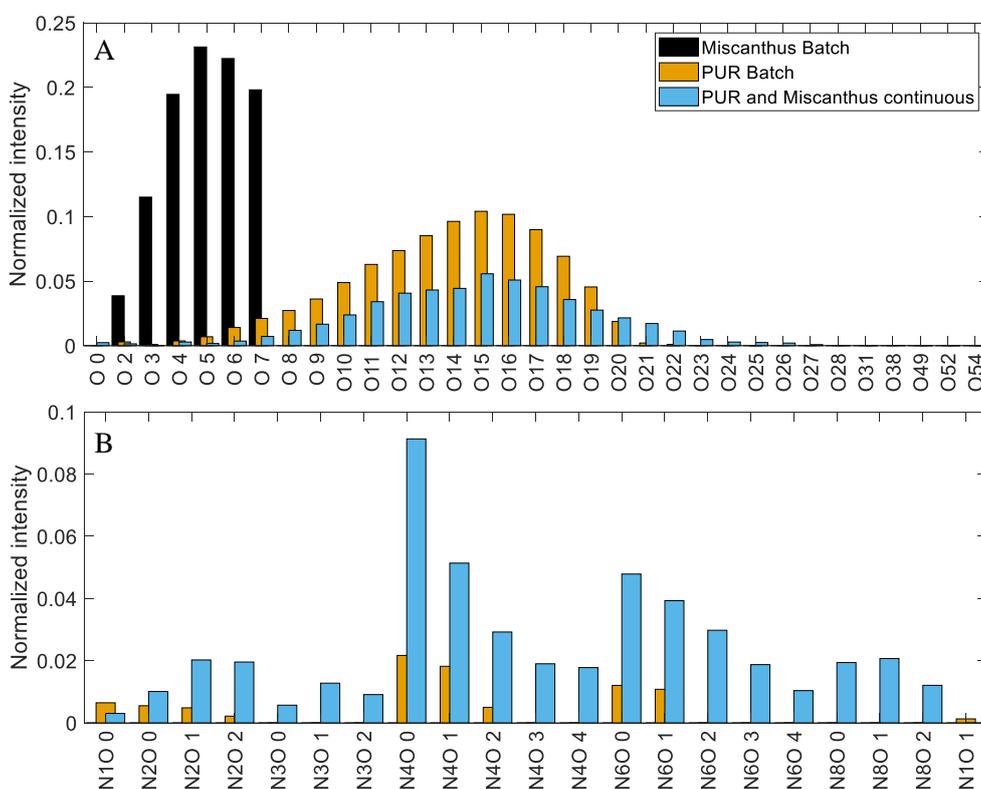

Figure 5 – Normalized intensity for all compounds in each $N_yO_x$ families for M batch biocrude (325 °C), PUR batch (325 °C) and continuous oil. N = 0 (A); N ≥ 1 (B)

In the PUR and M continuous oil, the most intense peaks according to their accurate mass values belong to the classes $C_{(5+n)}H_{(12+2n)}O_{(1+m)}$, $C_{(5+n)}H_{(10+2n)}O_{(1+m)}$ and, $C_{(5+n)}H_{(8+2n)}O_{(1+m)}$ – with n from 0 to 80 and m from 0 to 40. A constant gap of $C_3H_6O$ is found for the most intense species, being compatible with polyol (Polypropylenglycols) structures. When comparing these polyols via a $C_n$ over $O_x$ plot, as depicted in Figure 6A, it is possible to verify a linear relation that also shows monomer addition as the molecular difference between compounds. The polyols identified in the co-HTL oil (Figure 6B) are similar to the ones identified in the pure PUR oil (Figure 6A), however the former are present in relatively smaller quantities – also shown in Figure 5A.

Figure 5 A and B depicts that despite the polyols in the co-HTL oil are distributed similarly to pure PUR oil (but with smaller relative intensities), both oils differ greatly in the nitrogen-containing species. The continuous oil presents a higher occurrence of species with nitrogen, specially $N_4$, $N_6$ and $N_8$ classes (as shown also in figure S10). Interestingly, no compounds containing $N_5$, nor $N_7$ were identified both in the co-HTL or pure PUR oil. The oil elemental composition presented in Table S2 also corroborates these results, depicting low nitrogen content for M oil in comparison to co-HTL oil, which has similar compositions for both batch and continuous processing.

Figure S8 depicts the Van Krevelen diagram for all nitrogen species identified in pure PUR and co-HTL oil. All oils demonstrated consistently lower than 0.2 O/C ratio, while H/C mostly below 2, commonly with 1 as median. The comparison between oils demonstrates that the co-HTL oil is more complex, tending to have a greater number of nitrogen families than pure PUR oil. For all oils, compounds of the same family tend to group around a short range of O/C ratios, while a relatively greater ratios of H/C.

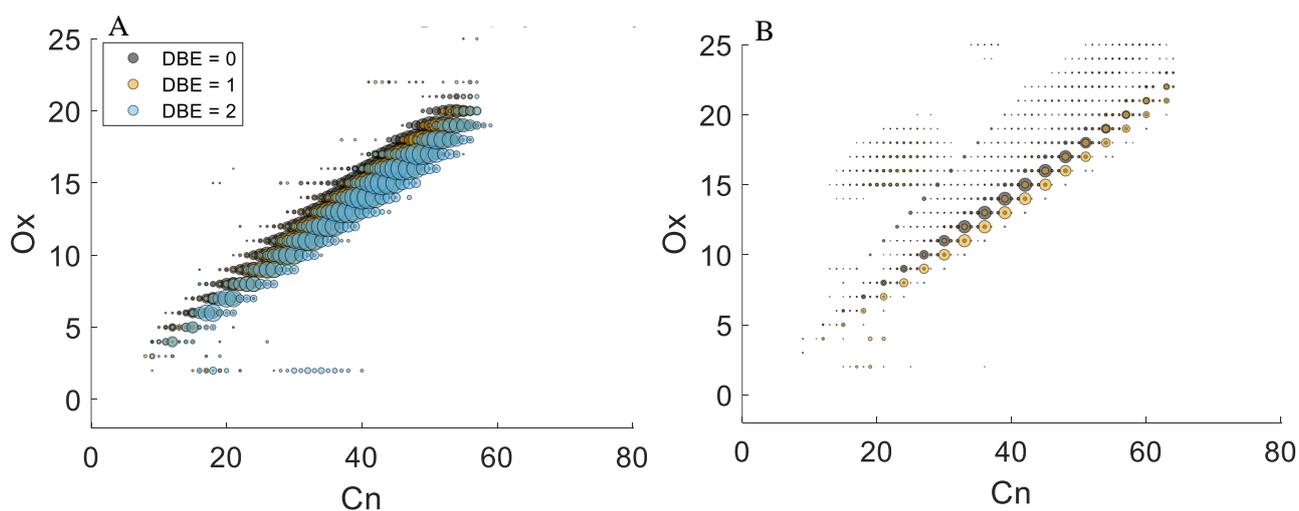

Figure 6 – Number of Carbon vs Oxygen atoms for $O_{x \geq 1}$ families (for DBE 0,1,2). The size of each dot is related to the relative abundance of the FTICR mass peak. PUR batch 325 °C (A); continuous oil (B)

The increase of O/C according to family maintaining similar H/C ratios indicates the molecular difference may be explained by the addition of hydroxyl side groups in these molecules. The atomic ratios of families that do not contain nitrogen are depicted in Figure 7, for batch PUR (A), M (B) and, continuous

(C) oils. Here, a very high H/C ratio for molecules with more than 1 oxygen atom is found both for PUR and the continuous oil in comparison to M biocrude. For both PUR-derived oils, a broad range of O/C ratios is identified, particularly in Figure 7C. The absence of compounds around H/C of 1 in Figure 7C thus indicates biomass-derived molecules rearrange in presence of PUR-derived compounds.

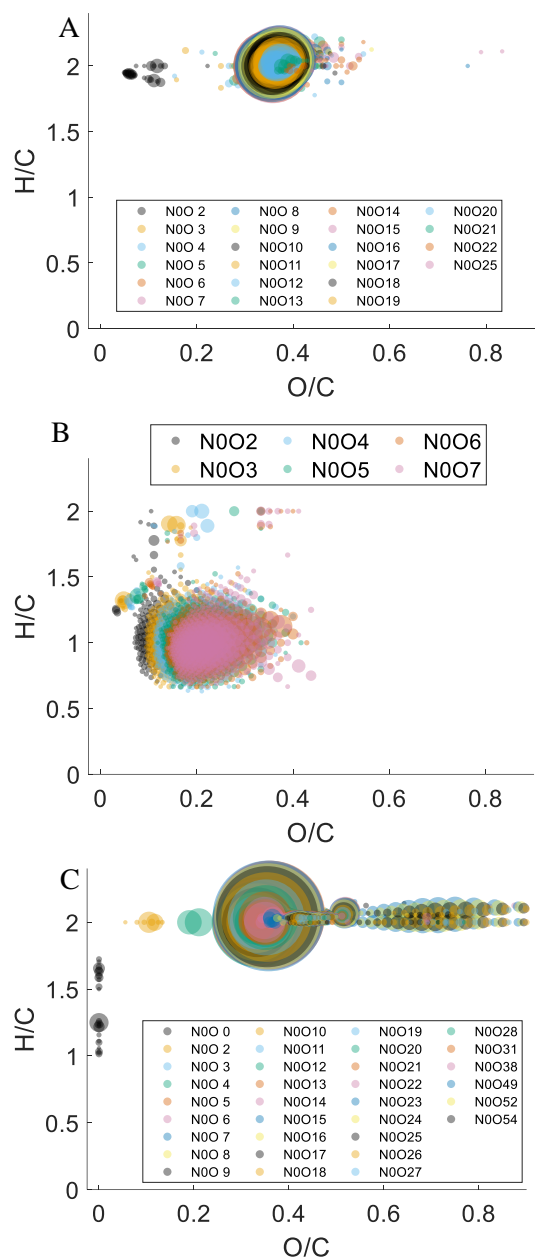

Figure 7 – Van Krevelen diagram for $O_x$ families. PUR batch 325 °C (A); M batch 325 °C (B) and; continuous oil (C)

A modified version of the Van Krevelen diagram, showing N/C ratio in the x-axis can be found in Figure S9. Here again, only the oils containing significant groups of N-containing structures are depicted. The difference between PUR batch and continuous oil indicates a clear recombination of molecules derived from M and PUR liquefaction, more specifically with a much greater variety of species with higher peak intensities, showing N/C ratios of up to 0.2. The greater number of nitrogen families is also depicted in Figure S10, where the DBE is plotted versus C number. PUR oil presents irregular distributions in DBE plots, very differently from the continuous oil. The latter contains classes that are grouped around an area in the $C_n$ vs DBE plot with species showing the highest relative intensities. These peaks vary according to family from 20 to 40 C number, generally increasing DBE with N and O contents. Such results are very indicative that the nitrogen containing species are connected to aromatic structures, which increase DBE drastically when comparing to biologically-derived HTL biocrude with high nitrogen content.[31]

In general, the complex oil derived from continuous processing of PUR and M is more aromatic than common biomass-derived biocrudes, still containing thousands of compounds. Furthermore, its composition includes molecules attached to polyol groups and short chain polyols as depicted in Figure 5 A. It is likely that the aromaticity favors the attachment amine groups in pair, given the families identified.

### 4.3. Overall reactions for high synergy effects

Maillard reactions are the main contributor to synergistic effects in hydrothermal liquefaction of biomass-derived protein in combination with sugars and other oxygen containing compounds.[29] Usually the recombination of monosaccharides and amino groups lead to the formation of pyrazines, indoles, pyrrolines, pyridines, pyrimidines and pyrazoles. Such heterocyclic compounds can also be a product of amino acids reacting in HTL media, but their formation is enhanced by the presence of sugars or other structures containing oxygen, for instance furans, ketones, organic acids, etc.[38]

Figure 8 depicts a possible reaction pathway for the observed high synergistic effect. Maillard reactions involving amines attached to aromatic groups occur faster than amino acids-like would[39], which can explain the high synergy effect observed in the co-HTL of PUR and M. The imine intermediates known in Maillard pathways[31] are formed and its consequent cyclisation increases aromaticity and stability for molecules. After, dehydration or hydrogenation reactions can follow leading to stable nitrogen heteroaromatics connected to alkyl side groups. However, a second reaction path can involve the still reactive sites present in the form of hydroxyl, ketone or acidic groups near hetero aromatic structures. In this case, recombination of these with other imine intermediates or amines adds of aromaticity and leads to more stable compounds with more hetero aromatic nitrogen atoms.

The nitrogen containing families described in section 4.2.3 depict a nitrogen atomic addition of 2, with major presence of $N_4$, $N_6$ and $N_8$ structures. This may indicate that imines or other compounds containing two nitrogen tend to recombine with the same family of intermediates, leading to very stable and highly aromatic structures. Overall, the reaction pathway proposed is very similar to biomass-derived nitrogen reactions, but as the amino groups are more active due to aromatic proximity, the synergy effects observed upon HTL processing are also more pronounced. Besides, the generation of compounds containing high levels of nitrogen (up to 8 atoms) differs from previously described biocrudes, despite similar oil elemental compositions.[29–31]

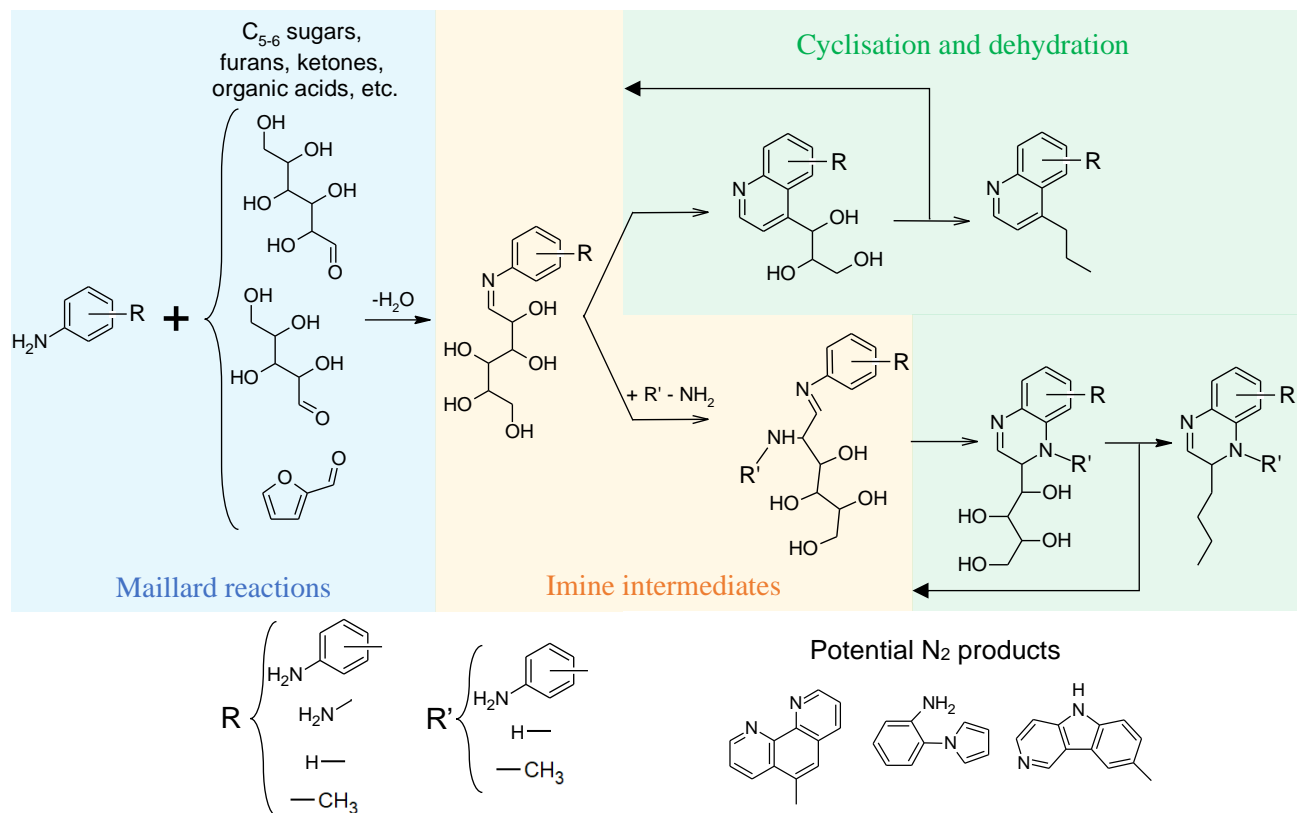
Figure 8 – Possible reaction pathway for high synergistic effect

## 5. CONCLUSION

The chemical recycling method presented can successfully enhance carbon and energy recovery in the form of an oil phase from lignocellulosics and PUR via combined HTL. From individual HTL carbon and energy recovery ranges of 35-45% and 35-55% respectively, the co-HTL models predicted in batch experiments a maximum of 70.8% for carbon and 75.0% for energy recovery. The model predictions were successfully validated via continuous HTL in a pilot plant operating at 316.4 °C, 55 kg.h$^{-1}$ slurry intake, which contained 78% of PUR and 22% of Miscanthus L., yielding 75% of energy and 71.2% carbon in the form of oil. Using the described pilot plant, a total energy efficiency of 60.7% was achieved (including utility requirements), leading to an EROI of 3.2.

The biocrude derived from co-HTL processing of PUR and lignocellulosic material was characterized as containing a significant portion of aromatic hydrocarbons with nitrogen heteroatoms, particularly with even numbers of those. Urethane bonds were fully hydrolyzed, generating primary amines, ethers, alcohol, aromatics and others, which recombined with biomass derived compounds into nitrogen containing species. Maillard reactions combining oxygen-containing molecules with amino groups attached to aromatics are probably the main reason for the superior performance observed in these trials in comparison to pure biomass HTL. The characteristics lead to formation of indoles and pyrazines, very stable compounds with high aromaticity which contribute to the maintenance of chemical energy in the form of oil product.

The results have demonstrated that HTL is a technically attractive method to recover carbon from synthetic resins aided by biomass. The combination of synthetic and biological materials can provide a new platform for circular economy by using the feedstock flexibility depicted here in favor of carbon recovery. The oil product depicted in this work can be considered a potential feedstock for refineries, which can be a pathway for including thermosetting resins in circular loops.

## ACKNOWLEDGEMENTS

The authors would like to thank Dan-Foam ApS (Tempur Denmark) for providing the sample for continuous flow processing. This project has received funding from the European Union's Horizon 2020 research and innovation grant agreement No. 764734 (HyFlexFuel − Hydrothermal Liquefaction: enhanced performance and feedstock flexibility for efficient biofuel production) and the Aarhus University Centre for Circular Bioeconomy (CBIO).

## REFERENCES

1. Akindoyo, J. O. *et al.* Polyurethane types, synthesis and applications-a review. *RSC Adv.* **6**, 114453–114482 (2016).

2. Nikje, M. M. A. *Recycling of Polyurethane Wastes*. (De Gruyter, 2019). doi:10.1515/9783110641592-201

3. Simón, D., Borreguero, A. M., de Lucas, A. & Rodríguez, J. F. Recycling of polyurethanes from laboratory to industry, a journey towards the sustainability. *Waste Manag.* **76**, 147–171 (2018).

4. Font, R., Fullana, A., Caballero, J. A., Candela, J. & García, A. Pyrolysis study of polyurethane. *J. Anal. Appl. Pyrolysis* **58–59**, 63–77 (2001).

5. Guo, X., Song, Z. & Zhang, W. Production of hydrogen-rich gas from waste rigid polyurethane foam via catalytic steam gasification. *Waste Manag. Res.* **38**, 802–811 (2020).

6. Zia, K. M., Bhatti, H. N. & Ahmad Bhatti, I. Methods for polyurethane and polyurethane composites, recycling and recovery: A review. *React. Funct. Polym.* **67**, 675–692 (2007).

7. Simón, D., García, M. T., De Lucas, A., Borreguero, A. M. & Rodríguez, J. F. Glycolysis of flexible polyurethane wastes using stannous octoate as the catalyst: Study on the influence of reaction parameters. *Polym. Degrad. Stab.* **98**, 144–149 (2013).

8. dos Passos, J. S., Glasius, M. & Biller, P. Screening of common synthetic polymers for depolymerization by subcritical hydrothermal liquefaction. *Process Saf. Environ. Prot.* **139**, 371–379 (2020).

9. Thomsen, L. *et al.* Hydrothermal liquefaction of sewage sludge; energy considerations and fate of micropollutants during pilot scale processing. *Water Res.* **183**, 116101 (2020).

10. Biller, P., Johannsen, I., dos Passos, J. S. & Ottosen, L. D. M. Primary sewage sludge filtration using biomass filter aids and subsequent hydrothermal co-liquefaction. *Water Res.* **130**, 58–68 (2018).

11. Faeth, J. L., Valdez, P. J. & Savage, P. E. Fast hydrothermal liquefaction of nannochloropsis sp. to produce biocrude. *Energy and Fuels* **27**, 1391–1398 (2013).

12. Conti, F. *et al.* Valorization of animal and human wastes through hydrothermal liquefaction for biocrude production and simultaneous recovery of nutrients. *Energy Convers. Manag.* **216**, 112925 (2020).

13. Aierzhati, A. *et al.* Experimental and model enhancement of food waste hydrothermal liquefaction with combined effects of biochemical composition and reaction conditions. *Bioresour. Technol.* **284**, 139–147 (2019).

14. Anastasakis, K., Biller, P., Madsen, R. B., Glasius, M. & Johannsen, I. Continuous Hydrothermal Liquefaction of Biomass in a Novel Pilot Plant with Heat Recovery and Hydraulic Oscillation. *Energies* **11**, 1–23 (2018).

15. Biller, P. & Ross, A. B. Potential yields and properties of oil from the hydrothermal liquefaction of microalgae with different biochemical content. *Bioresour. Technol.* **102**, 215–225 (2011).

16. Castello, D., Pedersen, T. H. & Rosendahl, L. A. Continuous hydrothermal liquefaction of biomass: A critical review. *Energies* **11**, (2018).

17. Xu, D. *et al.* Co-hydrothermal liquefaction of microalgae and sewage sludge in subcritical water: Ash effects on bio-oil production. *Renew. Energy* **138**, 1143–1151 (2019).

18. Raikova, S., Knowles, T. D. J., Allen, M. J. & Chuck, C. J. Co-liquefaction of Macroalgae with Common Marine Plastic Pollutants. *ACS Sustain. Chem. Eng.* **7**, 6769–6781 (2019).

19. Hongthong, S., Raikova, S., Leese, H. S. & Chuck, C. J. Co-processing of common plastics with pistachio hulls via hydrothermal liquefaction. *Waste Manag.* **102**, 351–361 (2020).

20. Hongthong, S., Leese, H. S. & Chuck, C. J. Valorizing Plastic-Contaminated Waste Streams through the Catalytic Hydrothermal Processing of Polypropylene with Lignocellulose. *ACS Omega* **5**, 20586–20598 (2020).

21. dos Passos, J. S., Glasius, M. & Biller, P. Hydrothermal Co-Liquefaction of Synthetic Polymers and Miscanthus giganteus : Synergistic and Antagonistic Effects. *ACS Sustain. Chem. Eng.* acssuschemeng.0c07317 (2020). doi:10.1021/acssuschemeng.0c07317

22. Biller, P. *et al.* Effect of hydrothermal liquefaction aqueous phase recycling on bio-crude yields and composition. *Bioresour. Technol.* **220**, 190–199 (2016).

23. Kozhinov, A. N., Zhurov, K. O. & Tsybin, Y. O. Iterative method for mass spectra recalibration via empirical estimation of the mass calibration function for fourier transform mass spectrometry-based petroleomics. *Anal. Chem.* **85**, 6437–6445 (2013).

24. Pellegrin, V. Molecular formulas of organic compounds: the nitrogen rule and degree of unsaturation. *J. Chem. Educ.* **60**, 626 (1983).

25. Gollakota, A. R. K., Kishore, N. & Gu, S. A review on hydrothermal liquefaction of biomass. *Renew. Sustain. Energy Rev.* **81**, 1378–1392 (2018).

26. Saba, A., Lopez, B., Lynam, J. G. & Reza, M. T. Hydrothermal Liquefaction of Loblolly Pine: Effects of Various Wastes on Produced Biocrude. *ACS Omega* **3**, 3051–3059 (2018).

27. Qian, L., Wang, S. & Savage, P. E. Fast and isothermal hydrothermal liquefaction of sludge at different severities: Reaction products, pathways, and kinetics. *Appl. Energy* **260**, 114312 (2020).

28. Elliott, D. C., Biller, P., Ross, A. B., Schmidt, A. J. & Jones, S. B. Hydrothermal liquefaction of biomass: Developments from batch to continuous process. *Bioresour. Technol.* **178**, 147–156 (2015).

29. Leng, L. *et al.* Nitrogen in bio-oil produced from hydrothermal liquefaction of biomass: A review. *Chem. Eng. J.* **401**, 126030 (2020).

30. Wu, B., Berg, S. M., Remucal, C. K. & Strathmann, T. J. Evolution of N-Containing Compounds

during Hydrothermal Liquefaction of Sewage Sludge. *ACS Sustain. Chem. Eng.* **8**, 18303–18313 (2020).

31. Lu, J., Li, H., Zhang, Y. & Liu, Z. Nitrogen Migration and Transformation during Hydrothermal Liquefaction of Livestock Manures. *ACS Sustain. Chem. Eng.* **6**, 13570–13578 (2018).

32. Lattimer, R. P., Polce, M. J. & Wesdemiotis, C. MALDI-MS analysis of pyrolysis products from a segmented polyurethane. *J. Anal. Appl. Pyrolysis* **48**, 1–15 (1998).

33. Esperanza, M. M., García, A. N., Font, R. & Conesa, J. A. Pyrolysis of varnish wastes based on a polyurethane. *J. Anal. Appl. Pyrolysis* **52**, 151–166 (1999).

34. Madsen, R. B., Anastasakis, K., Biller, P. & Glasius, M. Rapid Determination of Water, Total Acid Number, and Phenolic Content in Bio-Crude from Hydrothermal Liquefaction of Biomass using FT-IR. *Energy and Fuels* **32**, 7660–7669 (2018).

35. Pretsch, E., Bühlmann, P. & Badertscher, M. *Structure Determination of Organic Compounds*. (Springer Berlin Heidelberg, 2009). doi:10.1007/978-3-540-93810-1

36. Sudasinghe, N. *et al.* High resolution FT-ICR mass spectral analysis of bio-oil and residual water soluble organics produced by hydrothermal liquefaction of the marine microalga Nannochloropsis salina. *Fuel* **119**, 47–56 (2014).

37. Jarvis, J. M. *et al.* FT-ICR MS analysis of blended pine-microalgae feedstock HTL biocrudes. *Fuel* **216**, 341–348 (2018).

38. Zhang, C., Tang, X., Sheng, L. & Yang, X. Enhancing the performance of Co-hydrothermal liquefaction for mixed algae strains by the Maillard reaction. *Green Chem.* **18**, 2542–2553 (2016).

39. Brotzel, F., Ying, C. C., Mayr, H., Chu, Y. C. & Mayr, H. Nucleophilicities of primary and secondary amines in water. *J. Org. Chem.* **72**, 3679–3688 (2007).